\documentstyle[12pt]{article}
\catcode`\@=11

\font\tenmsa=msam10
\font\sevenmsa=msam7
\font\fivemsa=msam5
\font\tenmsb=msbm10
\font\sevenmsb=msbm7
\font\fivemsb=msbm5
\newfam\msafam
\newfam\msbfam
\textfont\msafam=\tenmsa  \scriptfont\msafam=\sevenmsa
  \scriptscriptfont\msafam=\fivemsa
\textfont\msbfam=\tenmsb  \scriptfont\msbfam=\sevenmsb
  \scriptscriptfont\msbfam=\fivemsb

\def\hexnumber@#1{\ifnum#1<10 \number#1\else
 \ifnum#1=10 A\else\ifnum#1=11 B\else\ifnum#1=12 C\else
 \ifnum#1=13 D\else\ifnum#1=14 E\else\ifnum#1=15 F\fi\fi\fi\fi\fi\fi\fi}

\def\msa@{\hexnumber@\msafam}
\def\msb@{\hexnumber@\msbfam}
\mathchardef\boxdot="2\msa@00
\mathchardef\boxplus="2\msa@01
\mathchardef\boxtimes="2\msa@02
\mathchardef\square="0\msa@03
\mathchardef\blacksquare="0\msa@04
\mathchardef\centerdot="2\msa@05
\mathchardef\lozenge="0\msa@06
\mathchardef\blacklozenge="0\msa@07
\mathchardef\circlearrowright="3\msa@08
\mathchardef\circlearrowleft="3\msa@09
\mathchardef\rightleftharpoons="3\msa@0A
\mathchardef\leftrightharpoons="3\msa@0B
\mathchardef\boxminus="2\msa@0C
\mathchardef\Vdash="3\msa@0D
\mathchardef\Vvdash="3\msa@0E
\mathchardef\vDash="3\msa@0F
\mathchardef\twoheadrightarrow="3\msa@10
\mathchardef\twoheadleftarrow="3\msa@11
\mathchardef\leftleftarrows="3\msa@12
\mathchardef\rightrightarrows="3\msa@13
\mathchardef\upuparrows="3\msa@14
\mathchardef\downdownarrows="3\msa@15
\mathchardef\upharpoonright="3\msa@16

\mathchardef\downharpoonright="3\msa@17
\mathchardef\upharpoonleft="3\msa@18
\mathchardef\downharpoonleft="3\msa@19
\mathchardef\rightarrowtail="3\msa@1A
\mathchardef\leftarrowtail="3\msa@1B
\mathchardef\leftrightarrows="3\msa@1C
\mathchardef\rightleftarrows="3\msa@1D
\mathchardef\Lsh="3\msa@1E
\mathchardef\Rsh="3\msa@1F
\mathchardef\rightsquigarrow="3\msa@20
\mathchardef\leftrightsquigarrow="3\msa@21
\mathchardef\looparrowleft="3\msa@22
\mathchardef\looparrowright="3\msa@23
\mathchardef\circeq="3\msa@24
\mathchardef\succsim="3\msa@25
\mathchardef\gtrsim="3\msa@26
\mathchardef\gtrapprox="3\msa@27
\mathchardef\multimap="3\msa@28
\mathchardef\therefore="3\msa@29
\mathchardef\because="3\msa@2A
\mathchardef\doteqdot="3\msa@2B

\mathchardef\triangleq="3\msa@2C
\mathchardef\precsim="3\msa@2D
\mathchardef\lesssim="3\msa@2E
\mathchardef\lessapprox="3\msa@2F
\mathchardef\eqslantless="3\msa@30
\mathchardef\eqslantgtr="3\msa@31
\mathchardef\curlyeqprec="3\msa@32
\mathchardef\curlyeqsucc="3\msa@33
\mathchardef\preccurlyeq="3\msa@34
\mathchardef\leqq="3\msa@35
\mathchardef\leqslant="3\msa@36
\mathchardef\lessgtr="3\msa@37
\mathchardef\backprime="0\msa@38
\mathchardef\risingdotseq="3\msa@3A
\mathchardef\fallingdotseq="3\msa@3B
\mathchardef\succcurlyeq="3\msa@3C
\mathchardef\geqq="3\msa@3D
\mathchardef\geqslant="3\msa@3E
\mathchardef\gtrless="3\msa@3F
\mathchardef\sqsubset="3\msa@40
\mathchardef\sqsupset="3\msa@41
\mathchardef\trianglerighteq="3\msa@44
\mathchardef\trianglelefteq="3\msa@45
\mathchardef\bigstar="0\msa@46
\mathchardef\between="3\msa@47
\mathchardef\blacktriangledown="0\msa@48
\mathchardef\blacktriangleright="3\msa@49
\mathchardef\blacktriangleleft="3\msa@4A
\mathchardef\blacktriangle="0\msa@4E
\mathchardef\triangledown="0\msa@4F
\mathchardef\eqcirc="3\msa@50
\mathchardef\lesseqgtr="3\msa@51
\mathchardef\gtreqless="3\msa@52
\mathchardef\lesseqqgtr="3\msa@53
\mathchardef\gtreqqless="3\msa@54
\mathchardef\Rrightarrow="3\msa@56
\mathchardef\Lleftarrow="3\msa@57
\mathchardef\veebar="2\msa@59
\mathchardef\barwedge="2\msa@5A
\mathchardef\doublebarwedge="2\msa@5B
\mathchardef\angle="0\msa@5C
\mathchardef\measuredangle="0\msa@5D
\mathchardef\sphericalangle="0\msa@5E
\mathchardef\varpropto="3\msa@5F
\mathchardef\smallsmile="3\msa@60
\mathchardef\smallfrown="3\msa@61
\mathchardef\Subset="3\msa@62
\mathchardef\Supset="3\msa@63
\mathchardef\Cup="2\msa@64

\mathchardef\Cap="2\msa@65

\mathchardef\curlywedge="2\msa@66
\mathchardef\curlyvee="2\msa@67
\mathchardef\leftthreetimes="2\msa@68
\mathchardef\rightthreetimes="2\msa@69
\mathchardef\subseteqq="3\msa@6A
\mathchardef\supseteqq="3\msa@6B
\mathchardef\bumpeq="3\msa@6C
\mathchardef\Bumpeq="3\msa@6D
\mathchardef\lll="3\msa@6E

\mathchardef\ggg="3\msa@6F

\mathchardef\circledS="0\msa@73
\mathchardef\pitchfork="3\msa@74
\mathchardef\dotplus="2\msa@75
\mathchardef\backsim="3\msa@76
\mathchardef\backsimeq="3\msa@77
\mathchardef\complement="0\msa@7B
\mathchardef\intercal="2\msa@7C
\mathchardef\circledcirc="2\msa@7D
\mathchardef\circledast="2\msa@7E
\mathchardef\circleddash="2\msa@7F
\def\ulcorner{\delimiter"4\msa@70\msa@70 }
\def\urcorner{\delimiter"5\msa@71\msa@71 }
\def\llcorner{\delimiter"4\msa@78\msa@78 }
\def\lrcorner{\delimiter"5\msa@79\msa@79 }
\def\yen{\mathhexbox\msa@55 }
\def\checkmark{\mathhexbox\msa@58 }
\def\circledR{\mathhexbox\msa@72 }
\def\maltese{\mathhexbox\msa@7A }
\mathchardef\lvertneqq="3\msb@00
\mathchardef\gvertneqq="3\msb@01
\mathchardef\nleq="3\msb@02
\mathchardef\ngeq="3\msb@03
\mathchardef\nless="3\msb@04
\mathchardef\ngtr="3\msb@05
\mathchardef\nprec="3\msb@06
\mathchardef\nsucc="3\msb@07
\mathchardef\lneqq="3\msb@08
\mathchardef\gneqq="3\msb@09
\mathchardef\nleqslant="3\msb@0A
\mathchardef\ngeqslant="3\msb@0B
\mathchardef\lneq="3\msb@0C
\mathchardef\gneq="3\msb@0D
\mathchardef\npreceq="3\msb@0E
\mathchardef\nsucceq="3\msb@0F
\mathchardef\precnsim="3\msb@10
\mathchardef\succnsim="3\msb@11
\mathchardef\lnsim="3\msb@12
\mathchardef\gnsim="3\msb@13
\mathchardef\nleqq="3\msb@14
\mathchardef\ngeqq="3\msb@15
\mathchardef\precneqq="3\msb@16
\mathchardef\succneqq="3\msb@17
\mathchardef\precnapprox="3\msb@18
\mathchardef\succnapprox="3\msb@19
\mathchardef\lnapprox="3\msb@1A
\mathchardef\gnapprox="3\msb@1B
\mathchardef\nsim="3\msb@1C
\mathchardef\napprox="3\msb@1D
\mathchardef\nsubseteqq="3\msb@22
\mathchardef\nsupseteqq="3\msb@23
\mathchardef\subsetneqq="3\msb@24
\mathchardef\supsetneqq="3\msb@25
\mathchardef\subsetneq="3\msb@28
\mathchardef\supsetneq="3\msb@29
\mathchardef\nsubseteq="3\msb@2A
\mathchardef\nsupseteq="3\msb@2B
\mathchardef\nparallel="3\msb@2C
\mathchardef\nmid="3\msb@2D
\mathchardef\nshortmid="3\msb@2E
\mathchardef\nshortparallel="3\msb@2F
\mathchardef\nvdash="3\msb@30
\mathchardef\nVdash="3\msb@31
\mathchardef\nvDash="3\msb@32
\mathchardef\nVDash="3\msb@33
\mathchardef\ntrianglerighteq="3\msb@34
\mathchardef\ntrianglelefteq="3\msb@35
\mathchardef\ntriangleleft="3\msb@36
\mathchardef\ntriangleright="3\msb@37
\mathchardef\nleftarrow="3\msb@38
\mathchardef\nrightarrow="3\msb@39
\mathchardef\nLeftarrow="3\msb@3A
\mathchardef\nRightarrow="3\msb@3B
\mathchardef\nLeftrightarrow="3\msb@3C
\mathchardef\nleftrightarrow="3\msb@3D
\mathchardef\divideontimes="2\msb@3E
\mathchardef\varnothing="0\msb@3F
\mathchardef\nexists="0\msb@40
\mathchardef\mho="0\msb@66
\mathchardef\thorn="0\msb@67
\mathchardef\beth="0\msb@69
\mathchardef\gimel="0\msb@6A
\mathchardef\daleth="0\msb@6B
\mathchardef\lessdot="3\msb@6C
\mathchardef\gtrdot="3\msb@6D
\mathchardef\ltimes="2\msb@6E
\mathchardef\rtimes="2\msb@6F
\mathchardef\shortmid="3\msb@70
\mathchardef\shortparallel="3\msb@71
\mathchardef\smallsetminus="2\msb@72
\mathchardef\thicksim="3\msb@73
\mathchardef\thickapprox="3\msb@74
\mathchardef\approxeq="3\msb@75
\mathchardef\succapprox="3\msb@76
\mathchardef\precapprox="3\msb@77
\mathchardef\curvearrowleft="3\msb@78
\mathchardef\curvearrowright="3\msb@79
\mathchardef\digamma="0\msb@7A
\mathchardef\varkappa="0\msb@7B
\mathchardef\hslash="0\msb@7D
\mathchardef\hbar="0\msb@7E
\mathchardef\backepsilon="3\msb@7F
\def\Bbb{\ifmmode\let\next\Bbb@\else
 \def\next{\errmessage{Use \string\Bbb\space only in math mode}}\fi\next}
\def\Bbb@#1{{\Bbb@@{#1}}}
\def\Bbb@@#1{\fam\msbfam#1}

\catcode`\@=\active

\def\N{{\Bbb N}}

\def\<{\langle}
\def\>{\rangle}

\def\cobicross{{\triangleright\!\!\!\blacktriangleleft}}
\def\bicross{{\blacktriangleright\!\!\!\triangleleft}}

\def\tens{\mathop{\otimes}}

\def\s#1{{}_{\scriptscriptstyle(#1)}}
\def\o{{}_{\scriptscriptstyle(1)}}
\def\t{{}_{\scriptscriptstyle(2)}}

\def\so{{{}_{\scriptscriptstyle[1]}}}

\def\note#1{}

\def\equad{\kern -1.7em}

\def\1{\hbox{\bf 1}}
\def\k{\kappa}
\def\s{\sigma}
\def\U{{\cal U}}
\def\trr{\triangleright}

\def\ftt{\footnotetext}
\def\ftm{\footnotemark}
\title{$\kappa$-Deformation of Poincar\'e Superalgebra with Classical
            Lorentz Subalgebra and its Graded Bicrossproduct Structure}

\author{\em P.\ Kosi{\'n}ski \ftm[1]  \ftm[5]  ,
            J.\ Lukierski \ftm[2]  \ftm[6]  ,
            P.\ Ma{\'s}lanka \ftm[3]  \ftm[5],
             and J.\ Sobczyk \ftm[4]  \ftm[6]
}
\date{}
\def\ccop#1{#1 \tens {\bf 1} + {\bf 1} \tens #1}
\def\<{\left<}
\def\>{\right>}
\def\pair#1#2{\<\right.#1,\,#2\left.\>}
\def\Np{N^{(+)}}
\def\Qp{Q^{(+)}}
\def\e{\epsilon}
\def\ben{\begin{enumerate}}
\def\een{\end{enumerate}}
\def\:{\,\,:\,\,\,}

\def\TK#1{T_{4;2}^{\kappa(#1)}}

\def\da{\hat\alpha}
\def\ha#1{\widehat\alpha^{(#1)}}
\def\hb#1{\widehat\beta^{(#1)}}
\def\to#1{\,\,{\stackrel{#1}\longrightarrow}\,\,}
\def\Pc{{\cal P}_4}
\def\({\left(}
\def\){\right)}
\def\tens{\otimes}
\def\~{\widetilde}
\def\dsp{\displaystyle}
\def\poin{Poincar\'e }

\def\bowti{\triangleright\!\!\!<}

\def\btc{\,\cobicross\,}
\def\ctb{\,\bicross\,}
\newcounter{popnr}
\renewcommand{\theequation}{\arabic{section}.\arabic{equation}}
\def\alpheqn{\setcounter{popnr}{\value{equation}}
             \stepcounter{popnr}
             \setcounter{equation}{0}
             \def\theequation{\arabic{section}.\arabic{popnr}\alph{equation}}
             }
\def\reseteqn{\setcounter{equation}{\value{popnr}}
              \def\theequation{\arabic{section}.\arabic{equation}}
              }
\newcommand{\beq}{\begin{eqnarray}}
\newcommand{\eeq}{\end{eqnarray}}
\newcommand{\beqq}{\begin{eqnarray*}}
\newcommand{\eeqq}{\end{eqnarray*}}

\def\bel#1{\begin{equation}\label{#1}}
\def\be{\begin{equation}}
\def\ee{\end{equation}}
\def\o{\overline}
\def\sPk{{\cal P}^\k_{4;1}}
\def\sP{{\cal P}_{4;1}}
\def\sT{T^\k_{4;2}}
\def\m{\mu}
\def\n{\nu}
\def\ro{\rho}
\def\r#1{(\ref{#1})}
\def\t{\tau}
\def\Mo#1#2{M^{(0)}_{#1#2}}
\def\N#1{N^{(#1)}}
\def\P#1{P^{(#1)}}

\def\Qo#1{Q^{(0)}_{#1}}
\def\bl{\alpheqn}
\def\el{\reseteqn}
\def\a{\alpha}
\def\b{\beta}
\def\ka{{\dot\a}}
\def\kb{{\dot\b}}
\def\ba{\begin{array}}
\def\ea{\end{array}}
\def\cop{\Delta}
\def\epk#1{{\dsp e^{#1\frac{P_0}\k}}}
\def\epkk#1{{\dsp e^{#1\frac{P_0}{2\k}}}}
\def\epkc#1{{\dsp e^{#1\frac{P_0}{4\k}}}}
\def\so{O(1,3;2)}
\def\kdef{$\k$-deformed }
\def\0{^{(0)}}
\def\po#1{\frac\partial{\partial #1}}
\def\vP{\vec P}
\def\vM{\vec M}

\begin{document}
\maketitle
\thispagestyle{empty}
\begin{abstract}
The $\k$-deformed $D=4$ Poincar{\'e} superalgebra  written in
Hopf superalgebra form is transformed to the basis with classical
Lorentz subalgebra generators.
We show that in such a basis the $\k$-deformed $D=4$ \poin superalgebra
can be written as graded bicrossproduct. We show that the $\k$-deformed
$D=4$ superalgebra acts covariantly on $\k$-deformed chiral superspace.
\end{abstract}

\def\thefootnote{\fnsymbol{footnote}}

\ftt[1]{Institute of Physics, University of
{\L}\'od\'z, ul. Pomorska 149/153, 90-236 {\L}\'od\'z, Poland.}
\ftt[2]{SISSA, via Beirut 9, Trieste-Miramare, Italy, on leave of absence from
the
Institute for Theoretical Physics, University of Wroc{\l}aw,
pl. Maxa Borna 9, 50-204 Wroc{\l}aw, Poland.}
\ftt[3]{Dept. of
Functional Analysis, Institute of Mathematics, University of {\L}{\'o}d{\'z},
ul. S. Banacha 22, 90-238  {\L}\'od\'z,
 Poland.}
\ftt[4]{International Centre for Theoretical Physics, 34100 Trieste,
Italy,
on leave of absence from Institute for Theoretical Physics, University
of Wroc{\l}aw, pl. Maxa Borna 9, 50-204 Wroc{\l}aw, Poland.}
\ftt[5]{Partially
supported by KBN grant 2P 302 21706.}
\ftt[6]{Partially supported by KBN grant 2P 302 08706.}
\setcounter{footnote}{0}
\def\thefootnote{\arabic{footnote}}

\newpage \setcounter{page}{1}

\section{Introduction}

Following the formulation of $D=4$ $\kappa$-deformed Poincar{\'e}
algebra [1--3]
recently also the $\kappa$-deformation of $D=4$ $N=1$ Poincar{\'e}
superalgebra was given \cite{4}. Both deformations were firstly obtained in the
framework
of Hopf (super)algebras by the quantum contraction procedure of ${\cal
U}_q(O(3,2))$ and ${\cal U}_q(OSp(1|4))$ and have
the following properties:
\begin{itemize}
\item[a)] The fourmomenta remain commutative, but noncocommutative,
\item[b)] The three-dimensional rotations remain classical as Hopf
algebras,
\item[c)] The Lorentz generators do not form a subalgebra (neither the
Lie subalgebra nor the Hopf subalgebra).
\end{itemize}
The first two properties imply that the deformation is "mild", and does
not affect the rotational symmetry of nonrelativistic physics. The
property c) is not convenient from physical point of view -- in
particular there are difficulties with the interpretation of finite
$\kappa$--deformed Lorentz transformations which do not form a Lie group
\cite{5}. Recently. however , it has been given by Majid and Ruegg
\cite{6} the
basis of quantum $\kappa$-Poincar{\'e} algebra, describing it
as a bicrossproduct of classical Lorentz Hopf algebra $O(3,1)$ with the
Hopf algebra $T_4^\k$ of commuting fourmomenta equipped with
$\k$-deformed coproduct
\bel{1.1}
\Pc^\k = O(1,3) \btc T_4^\k\,.
\ee
In such a framework the
classical Lorentz algebra $O(1,3)$ (but not a classical Hopf algebra
$O(1,3)$ !) is the subalgebra of ${\cal P}_4^\k$, and $T^\k_4$ forms a
Hopf subalgebra of ${\cal P}_4^\k$.

The aim of this paper is to find analogous basis for $\kappa$-deformed
Poincar{\'e} superalgebra, with classical Lorentz subalgebra and
commuting fourmomenta, which supersymmetrize the Majid--Ruegg
bicrossproduct basis for
$\kappa$--Poincar{\'e} algebra\footnote{By supersymmetrization of
$\kappa$--Poincar{\'e} algebra ${\cal P}_4^\k$ we mean the
$\k$-Poincar{\'e} superalgebra $\sPk$  which
after formally putting in the bosonic sector the supercharges equal to zero
reduces to ${\cal P}^\k_4$.}.
Such a formulation is derived (see Sect. 2) by
nonlinear change of the basis of $\kappa$-Poincar{\'e} superalgebra ,
obtained previously in \cite{4} from the quantum contraction of ${\cal U}_q
(OSp (1;4))$. It appears that the $\kappa$-Poincar{\'e} superalgebra
${\cal P}^\k_{4;1}$ can be written e.g.\
as the following graded bicrossproduct
\footnote{We give here only one possibility -- in fact there are four
ways of expressing $\sPk$ as graded bicrossproduct (see Sect. 3).}
which extends supersymmetrically the formula (1.1):
\bel{1.2}
\sPk = O(1,3;2) \btc  T^\k_{4;2}\,,
\ee
where $O(1,3;2)$ is the classical superextension of the Lorentz
algebra:\\
($\eta_{\mu\nu}= \hbox{diag}(1,-1,-1,-1)
$)\footnote{ We denote by $I^{(0)}_A$ the generators of classical Lie
Hopf (super)algebras, with primitive coproducts $\cop (I\0_A)= 1  \tens
I\0_A + I\0_A \tens 1$.}.
\bel{1.3}
[\Mo\m\n,\Mo\ro\t]= i (\eta_{\m\t}\Mo\n\ro + \eta_{\n\ro}\Mo\m\t-
\eta_{\m\ro}\Mo\n\t-\eta_{\n\t}\Mo\m\ro)
\ee
with two complex supercharges $Q_\a$ ($\a=1,2$) satisfying the relations
\bl
\bel{1.4a}
{}[\Mo\m\n,\Qo\a]=\frac i 2 (\s_{\m\n})_\a{}^\b\Qo\b\,,\qquad
\{\Qo\a,\Qo\b\}=0\,,
\ee
and $\o\sT$ describes the complex Hopf superalgebra ($\m,\n =
0,1,2,3$)
\bel{1.4.b}
\{\o Q_\ka,\o Q_\kb\}=0\,,\qquad [\o Q_\ka, P_\m]=[P_\m,P_\n]=0\,,
\ee
\el
supplemented by the following coproducts
\bel{1.5}
\ba{rcl}
\cop P_i& =& \epk- \tens P_i + P_i \tens \1\,,\\[1mm]
\cop P_0& =& P_0 \tens \1 + \1 \tens P_0\,,\\[1mm]
\cop\o  Q_\ka &=& \o Q_\ka\tens  \epkk-  + \o Q_\ka \tens \1\,.
\ea
\ee
In Sect.\ 3  we shall show that the $\k$-deformation of $N=1$ \poin
superalgebra can be described by the $\k$-dependent action of the
superalgebra $O(1,3;2)$ on $\o\sT$, modifying in the algebraic sector
the classical $O(1,3;2)$-covariance relations for the $\o\sT$
generators,
as well as $\k$-dependent coaction of $\o\sT$ on $O(1,3;2)$, modifying the
classical coproducts of the $\so$ generators.

The bicrossproduct structure of $\sP^\k$ implies that the dual Hopf algebra
$(\sP^\k)^*$ describing quantum $N=1$ \poin supergroup has also the
bicrossproduct structure (see e.g.\ [8])
\bel{1.6}
(\sPk)^* = (\so)^* \ctb (\o\sT)^*\,,
\ee
where $(\o\sT)^*$ describes $\k$-deformed complex chiral superspace
$(x_\m,\theta_\ka)$  on which acts covariantly the
\kdef superalgebra $\sPk$. The \kdef superspace has been introduced
recently in \cite{7}, but only the bicrossproduct structure of $\sPk$
permits to show that its chiral part transforms covariantly under the
\kdef supersymmetry transformations. These transformations, obtained by
the relation \r{1.2} are described in Sect.\ 4. In such a way we
have all ingredients which are needed for the construction of \kdef
chiral superfields formalism, what we shall present in our next
publications.

\section{The  $\kappa$-deformed $N=1$ $D=4$ Poincar{\'e}
superalgebra with classical Lorentz generators.}
\setcounter{equation}{0}

Let us recall firstly the formulae describing the $\kappa$-deformed $N=1$ $D=4$
Poincar{\'e} superalgebra as the noncommutative  and noncocommutative
real Hopf superalgebra. We have the following set of relations
\cite{4}.
\begin{description}
\item[a)] Lorentz sector $(M_{\mu\nu} = (M_i, N_i)$, where $M_i =
\frac{1}{2}
\epsilon_{ijk}M_{jk}$ describe the non-relativistic $O(3)$ rotations,
and $N_i=M_{0i}$ describe boosts).
  \begin{description}
\item[$i)$] {\it algebra} ($\vP=(P_1,P_2,P_3)$, $\vM=(M_1,M_2,M_3)$)\\
\alpheqn
\beq
[M_i, M_j] = i\epsilon_{ijk}M_k\,, \qquad  [M_i, L_j] = i\epsilon_{ijk}L_k
\eeq
\beq 
\!\!\!\!\!\!\!\!\!\!\!\!\!\!\!
[L_i, L_j] = -i\epsilon_{ijk}\(M_k \cosh{{P_0}\over {\kappa}} - {1\over
{8\kappa}}T_k \sinh{{P_0}\over {2\kappa}} 
+ {1\over {16\kappa^2}}P_k (T_0 - 4\vP\vM)\)
\eeq
\reseteqn
where  $(\mu = 0, 1, 2, 3)$
\bel{2.2}
T_{\mu} = Q^A(\sigma_{\mu})_{A\dot B}Q^{\dot B}
\ee
\item[$ii)$] {\it coalgebra}\\
\alpheqn
\beq
\Delta (M_i) = M_i \otimes \hbox{\bf 1} + \hbox{\bf 1} \otimes M_i
\eeq
\beq
\!\!\!\!\!\!\!\!\!\!\!\begin{array}{rcl}
\Delta (L_i)& =& L_i \otimes e^{{P_0}\over {2\kappa}} + e^{-{{P_0}\over
{2\kappa}}}\otimes L_i 
+ {1\over {2\kappa}}\epsilon_{ijk} (P_j \otimes
  M_k e^{{P_0}\over {2\kappa}} + M_j e^{-{{P_0}\over
{2\kappa}}}\otimes P_k) \\[2mm]
&&+{i\over {8\kappa}}
(\sigma_i)_{\dot\alpha\beta}\(\overline{Q}_{\dot\alpha}e^{-{{P_0}\over
{4\kappa}}}\otimes Q_{\beta}e^{{P_0}\over {4\kappa}} +
Q_{\beta}e^{-{{P_0}\over {4\kappa}}}\otimes
\overline{Q}_{\dot\alpha}e^{{P_0}\over {4\kappa}}\)
\end{array}
\eeq
\reseteqn
\item[$iii)$] {\it antipodes}\\
\bel{2.4}
\begin{array}{rcl}
S(M_i)&=& - M_i\,,\\[2mm]
S(N_i)&=& - N_i + {{3i}\over {2\kappa}}P_i - {i\over
{8\kappa}}\(Q\sigma_i\overline{Q} + \overline{Q}\sigma_i Q\)\,,
\end{array}
\ee
\end{description}
\item[b)] Fourmomenta sector $P_{\mu} = (P_i, P_0)$
$(\m,\n=0,1,2,3)$
  \begin{description}
\item[$i)$] {\it algebra}\\
\alpheqn
\beq
\lbrack M_i, P_j\rbrack  = i\epsilon_{ijk}P_k\,,  \qquad \lbrack M_j,
P_0\rbrack  = 0\,,
\eeq
\vskip -0.3cm
\beq
\lbrack L_i, P_j\rbrack   = i\kappa\delta_{ij}\sinh{{P_0}\over
{\kappa}}\,,
\qquad
\lbrack L_i, P_0\rbrack   = iP_i\,,
\eeq
\beq
\lbrack P_{\mu}, P_{\nu}\rbrack  = 0\,,
\eeq
\reseteqn
\item[$ii)$] {\it coalgebra}\\
\alpheqn
\beq
\Delta (P_i)  = P_i \otimes e^{{P_0}\over {2\kappa}} + e^{-{{P_0}\over
{2\kappa}}} \otimes P_i\,,
\eeq
\beq
\Delta (P_0)  = P_0 \otimes \hbox{\bf 1} + \hbox{\bf 1} \otimes P_0\,,
\eeq
\reseteqn
  \end{description}
The antipode is given by the relation
$S(P_{\mu}) = -P_{\mu}$.
\item[c)] Supercharges sector \cite{4}
  \begin{description}
\item[$i)$] {\it algebra}
\alpheqn
\beq
\begin{array}{ll}
\{Q_{\alpha}, \o Q_{\dot\beta}\} & =
4\kappa\delta_{\alpha\beta}\sin{{P_0}\over {2\kappa}} - 2P_i
(\sigma_i)_{\alpha\dot\beta}\,, \\[2mm]
\{Q_{\alpha}, Q_{\beta}\} & = \{\o Q_{\dot\alpha},\o Q_{\dot\beta}\} =
0\,,
\end{array}
\eeq
\beq
\lbrack M_i, Q_{\alpha}\rbrack = -{1\over 2} (\sigma _i)_{\alpha}^{\beta}
Q_{\beta}\,, \quad
\lbrack M_i,\o Q_{\dot\alpha}\rbrack = {1\over 2} (\sigma
_i)_{\dot\alpha}^{\dot\beta}\o Q_{\dot\beta}\,,
\eeq
\beq
\lbrack L_i, Q_{\alpha}\rbrack = -{i\over 2}\cosh{{P_0}\over {2\kappa}} (\sigma
_i)_{\alpha}^{\beta}Q_{\beta}\,,\quad
\lbrack L_i,\o Q_{\dot\alpha}\rbrack = {i\over 2}\cosh{{P_0}\over {2\kappa}}
(\sigma_i)_{\dot\alpha}^{\dot\beta}\o Q_{\dot\beta}\,,
\eeq
\beq
\begin{array}{ll}
\lbrack P_{\mu}, Q_{\alpha}\rbrack  & = \lbrack P_{\mu},\o
Q_{\dot\beta}\rbrack  = 0\,,
\end{array}
\eeq
\reseteqn
\item[$ii)$] {\it coalgebra}
\beq
\begin{array}{ll}
\Delta (Q_{\alpha}) &\dsp = Q_{\alpha} \otimes e^{{P_0}\over {4\kappa}} +
e^{-{{P_0}\over {4\kappa}}} \otimes Q_{\alpha}\,,\\[2mm]
\Delta (\o Q_{\dot\alpha}) &\dsp =\o  Q_{\dot\alpha} \otimes e^{{P_0}\over
{4\kappa}} +
e^{-{{P_0}\over {4\kappa}}} \otimes \o Q_{\dot\alpha}\,,
\end{array}
\eeq
\item[$iii)$] {\it antipodes}
\beq
S(Q_{\alpha}) = - Q_{\alpha}\,, \qquad S(\o Q_{\dot\alpha}) = - \o
Q_{\dot\alpha}\,.
\eeq
  \end{description}
\end{description}
On the basis of the relations (2.3) - (2.7) one can
single out the following features of the quantum superalgebra
${\cal U}_{\kappa} ({\cal P}_{4;1})$ :
\begin{description}
\item[$i)$] The algebra coproducts and antipodes of Lorentz boosts $N_i$
do depend on $Q_{\alpha},\o Q_{\dot\alpha}$ i.e. the $\kappa$-deformed
Poincar{\'e} as well as Lorentz sectors do not form the Hopf
subalgebras.
\item[$ii)$] Putting formally in the formulae (2.1) - (2.6) $Q_{\alpha} =
Q_{\dot\alpha} = 0$ one obtains the $\kappa$-deformed Poincar{\'e}
algebra considered in \cite{3}, i.e.
\beq
{\cal U}_{\kappa} ({\cal P}_{4;1}) \left |_{Q_{\alpha} = Q_{\dot\alpha} =
0}\right . = {\cal U}_{\kappa}({\cal P}_4)
\eeq
\end{description}
In order to remove from the formulae (2.1b) the supercharge-dependent
terms one can introduce the following two complex Lorentz boosts
\beq
L^{(\pm )}_i =L_i \pm \frac{i}{8\kappa}T_i
\eeq
complex--conjugated to each other ($(L^{(+)}_i )^{+} = L^{(-)}_i$). One
gets however
\beq
[L_i^{(\pm )},L_j^{(\pm )} ]=-i\epsilon_{ijk} (M_k \cosh
\frac{P_0}{\kappa} -\frac{1}{4\kappa^2}P_k (\vP\vM))
\eeq
Using (2.7 d) and (2.11) one obtains
\beq
[L^{(\pm )}_i ,P_j ] =i\kappa\delta_{ij} \sinh \frac{P_0}{\kappa}\,, \qquad
[L^{(\pm )}_i, P_0 ]=iP_i        .
\eeq
One can also calculate that
\beq 
\begin{array}{rl}
\Delta L^{(+)}_i =& L^{(+)}_i\otimes e^{\frac{P_0}{2\kappa}} +
e^{-\frac{P_0}{2\kappa}}\otimes L^{(+)}_i + \\[2mm]
&+\frac{1}{2\kappa}\epsilon_{ijk}(P_j \otimes M_k
e^{\frac{P_0}{2\kappa}}
+M_j e^{-\frac{P_0}{2\kappa}}\otimes P_k ) 
+\frac i{4\k} (\s_i)_{\ka\b} \epkc- \o Q_\ka \tens \epkc{} Q_\b\,,
\end{array}
\eeq
\beq 
\begin{array}{rl}
\Delta L^{(-)}_i =& L^{(-)}_i\otimes e^{\frac{P_0}{2\kappa}} +
e^{-\frac{P_0}{2\kappa}}\otimes L^{(-)}_i + \\[2mm]
&+\frac{1}{2\kappa}\epsilon_{ijk}(P_j \otimes M_k
e^{\frac{P_0}{2\kappa}}
+M_j e^{-\frac{P_0}{2\kappa}}\otimes P_k ) 
-\frac i{4\k} (\s_i)_{\ka\b} \epkc- Q_\b \tens \epkc{} \o Q_\ka\,,
\end{array}
\eeq

Further one can calculate the formulae in the supercharge sector. The
modification (2.11) of the boost operators leads to the following
covariance relations:
\alpheqn
\beq
[L^{(\pm )}_i,Q_{\alpha}]=-\frac{i}{2} e^{\pm\frac{P_0}{2\kappa}}
(\sigma_i Q)_{\alpha} \pm \frac{i}{4\kappa}P_i Q_{\alpha}\pm
\frac{1}{4\kappa}\epsilon^{ijk} P_k (\sigma_i Q)_{\alpha}
\eeq
\beq
[L^{(\pm )}_i, \overline{Q_{\dot\alpha}}]=-\frac{i}{2}
e^{\mp\frac{P_0}{2\kappa}}(\overline{Q} \sigma_i )_{\dot\alpha} \mp
\frac{i}{4\kappa} P_i \overline{Q}_{\dot\alpha}\pm
\frac{1}{4\kappa}\epsilon^{ijk}P_k (\overline{Q}\sigma_i
)_{\dot\alpha}
\eeq
\reseteqn
The relations (2.7a-b), (2.7d) and (2.16a-b) describe the
supersymmetric extensions of the $\kappa$-Poincar{\'e} algebra ${\cal
U}_{\kappa}({\cal P}_4)$, given in \cite{3}.

In order to obtain the basis describing the bicrossproduct structure
we introduce the following pairs of transformations
\bel{2.17}
\begin{array}{ll}
N^{(\pm )}_i&=\frac{1}{2}\{L^{(\pm
)}_i,e^{\mp\frac{P_0}{2\kappa}}\}\mp\frac{1}{2\kappa}\epsilon_{ijk} M_jP_k
e^{\mp\frac{P_0}{2\kappa}}\,,\\[2mm]
\P\pm_i&=P_ie^{\mp\frac{P_0}{2\kappa}}\,,
\end{array}
\ee
\bel{2.18}
Q^{(\pm )}_\a=e^{\mp\frac{P_0}{4\kappa}}Q_\a\,,\quad \quad \o{Q}_\ka^{(\pm
)}=e^{\pm\frac{P_0}{4\kappa}}\o Q_\ka\,.
\ee
It should be pointed out that for the generators $(\N+_i, \P+_i)$
the transformation \r{2.17} coincides with the one given in \cite{6}.
One obtains the following two Hopf superalgebra structures:\\
{\bf a)} Lorentz sector $(M_{\mu\nu}^{(\pm )}=(M_i,N_i^{(\pm )})$\\
{\it a1) algebra}\\
\beq
\begin{array}{ll}
\lbrack M_i,M_j\rbrack&=i\epsilon_{ijk}M_k\,,\\[2mm]
\lbrack M_i,N_k^{(\pm )}\rbrack &=i\epsilon_{ijk}N_k^{(\pm )}\,,\\[2mm]
\lbrack N^{(\pm )}_i ,N^{(\pm )}_j\rbrack &=-i\epsilon_{ijk}M_k\,,
\end{array}
\eeq
{\it a2) coalgebra}\\
\alpheqn
\beq
\Delta (M_i )&=&M_i\otimes {\bf 1}+{\bf 1}\otimes M_i\,,
\eeq
\beq
\begin{array}{rcl}
\Delta (N^{(+)}_i)&=&N^{(+)}_i \otimes {\bf
1}+e^{-\frac{P_0}{\kappa}}\otimes N^{(+)}_i +\null\\[2mm]
&&\null+\frac{1}{\kappa}\epsilon_{ijk}{P}^{(+)}_j\otimes M_k
+\frac{i}{4\kappa}{(\sigma_i)_{\ka\b}}e^{-\frac{P_0}{\kappa}}
\o {Q}^{(+)}_\ka \otimes Q^{(+)}_\b\,,
\end{array}
\eeq
\beq
\begin{array}{rcl}
\Delta (N^{(-)}_i)&=&N^{(-)}_i \otimes \epk{} +{\bf 1}
 \otimes N^{(-)}_i \\[2mm]
&&-\frac{1}{\kappa}\epsilon_{ijk}M_k \tens P^{(-)}_j
-\frac {i}{4\k}(\sigma_i)_{\ka\b}Q^{(-)}_\b \tens \epk{} \o {Q}^{(-)}_\ka \,,
\end{array}
\eeq
\el
{\it a3) antipode} ($T_i\equiv T_i^{(+)}=T_i^{(-)}$)\\
\bel{2.21}
\begin{array}{rcl}
S(M_i)&=&-M_i\,,\\[2mm]
S(N^{(\pm )}_i)&=&\dsp -\left[N_i^{(\pm)} \pm \frac 1 \k
\epsilon_{ijk}M_j P^{(\pm)}_k
\right.\\[2mm]
&&\left.\dsp \mp\frac{3i}{2\k} P_i^{(\pm)}  \pm \frac i {4\k}
\left(2\P\pm_i-T_i\epkk\mp
\right)\right]\epk\pm\,,
\end{array}
\ee
{\bf b)} Fourmomentum sector:
${P}^{(\pm)}_{\mu}=({P}^{(\pm)}_i,{P}^{(\pm)}_0=P_0)$\\
{\it b1) algebra}\\
\alpheqn
\beq
[\P\pm_{\mu},\P\pm_{\nu}]=0\,,
\eeq
\beq
[M_i,\P\pm_j]=i\epsilon_{ijk}\P\pm_k\,, \quad [M_i,P_0]=0\,,
\eeq
\beq
[N^{(\pm )}_i, \P\pm_j]
=\pm i\delta_{ij}\left[\frac{\k}{2}(1- e^{\mp\frac{2P_0}{\kappa}})
+\frac{1}{2\kappa} {\vec{P}{}^{(\pm)}}^2\right]\mp
\frac{1}{\kappa}\P\pm_i\P\pm_j\,,
\eeq
\beq
[N^{(\pm )}_i, P_0]=i\P\pm_i\,,
\eeq
\reseteqn
{\it b2) coalgebra}\\
\bel{2.23}
\Delta P_0=P_0\otimes {\bf 1}+{\bf 1}\otimes P_0\,,
\ee
\bel{2.24}
\begin{array}{ll}
\Delta \P+_i&=\P+_i\otimes{\bf
1}+e^{-\frac{P_0}{\kappa}}\otimes\P+_i\,, \\[2mm]
\Delta\P-_i&=\P-_i\otimes \epk{}
+{\bf 1} \otimes\P-_i\,,
\end{array}
\ee
{\it b3) antipode}\\
\beq
S(\P\pm_i)=-\epk\pm\P\pm_i\,,\qquad S(P_0)= -P_0\,,
\eeq


{\bf c)} Supercharge sector $(Q^{(\pm )}_{\alpha},\o {Q}^{(\pm
)}_{\alpha})$\\
{\it c1) algebra}
\beq
\begin{array}{ll}
\lbrack M_i,Q^{(\pm )}_{\alpha}\rbrack &=-\frac{1}{2}(\sigma_i Q^{(\pm
)})_{\alpha}\\[2mm]
\lbrack M_i, \o {Q}^{(\pm )}_{\ka}\rbrack &=\frac{1}{2}(\o {Q}^{(\pm )}\sigma_i
)_{\ka}
\end{array}
\eeq
\beq  
\begin{array}{ll}
\lbrack N^{(+)}_i,{Q}^{(+)}_{\alpha}\rbrack
&=-\frac{i}{2}(\sigma_i{Q}^{(+)})_\a \,,\\[2mm]
\lbrack N^{(+)}_i,\o {Q}^{(+)}_{\ka}\rbrack
&=-\frac{i}{2}e^{-\frac{P_0}{\kappa}}(\o {Q}^{(+)}\sigma_i)_{\ka}
+\frac{1}{2\kappa}\epsilon_{ikl}\P+_k(\o {Q}^{(+)}\sigma_l )_{\ka}\,,\\[2mm]
\lbrack N^{(-)}_i, {Q}^{(-)}_{\a}\rbrack
&=-\frac{i}{2}(\sigma_i {Q}^{(-)})_\a\,,\\[2mm]
\lbrack N^{(-)}_i,\o Q^{(-)}_{\ka}\rbrack
&=-\frac{i}{2}\epk{}
(\o Q^{(-)}\sigma_i)_{\ka}-\frac{1}{2\kappa}
\epsilon_{ikl}\P-_k(\o Q^{(-)}\sigma_l)_{\ka}\,,
\end{array}
\eeq
\beq
[Q^{(\pm )}_{\alpha}, \P\pm_{\mu}]=[\o {Q}^{(\pm )}_{\ka}, \P\pm_{\mu}]=0\,,
\eeq
\beq
\{Q^{(\pm)}_{\alpha}, \o
{Q}^{(\pm)}_{\dot\beta}\}=4\kappa\delta_{\alpha\dot\beta} \sinh
\(\frac{P_0}{2\kappa}\)-2e^{\pm\frac{P_0}{2\kappa}}\P\pm_i(\sigma_i
)_{\alpha\dot\beta}\,,
\eeq
{\it c2) coalgebra}\\
\beq
\Delta (Q^{(+)}_{\alpha})&=e^{-\frac{P_0}{2\kappa}} \tens Q^{(+)}_{\alpha}
+ Q^{(+)}_{\alpha}\tens {\bf 1}\,,\\[2mm]
\Delta (\o {Q}^{(+)}_{\alpha})&={\bf 1}\tens \o {Q}^{(+)}_{\alpha}
+\o {Q}^{(+)}_{\alpha}\tens e^{\frac{P_0}{2\kappa}}\,,\\[2mm]
\Delta (Q^{(-)}_{\alpha})&={\bf 1}\tens Q^{(-)}_{\alpha}
+ Q^{(-)}_{\alpha}\tens e^{\frac{P_0}{2\kappa}}\,,\\[2mm]
\Delta (\o {Q}^{(-)}_{\alpha})&=
e^{-\frac{P_0}{2\kappa}}\tens\o {Q}^{(-)}_{\alpha}
+\o {Q}^{(-)}_{\alpha}\tens{\bf 1}\,,
\eeq
{\it c3) antipodes}\\
\beq
S (Q^{(\pm )}_{\alpha})=-Q^{(\pm )}_{\alpha} \epkk\mp\,,\qquad
S(\o {Q}^{(\pm )}_{\alpha})=-\o {Q}^{(\pm )}_{\alpha}\epkk\mp\,,
\eeq
We see that we obtain two $\kappa$-deformed Poincar{\'e} Hopf superalgebras
(see also (2.15); $R=1,\ldots 14$)
\beq
{\cal U}_{\kappa}^{(\pm)} ({\cal P}_{4;1}):\~{I}^{(\pm
)}_R=(M_i,N_i^{(\pm)},\P\pm_i,P_0\,, Q^{(\pm )}_{\alpha},\o {Q}^{(\pm
)}_{\alpha})\,,\qquad \~I_R^{(+)} =\(\~I_R^{(-)}\)^\oplus\,.
\eeq
related by the nonstandard involution $\oplus$ (see also \cite{8})
introduced by the change of sign of $\k$ i.e.\
satisfying the following properties:
\bel{2.36}
\ba{c}
(ab)^\oplus = a^\oplus b^\oplus\,,\qquad \k^\oplus=-\k\,,\\[2mm]
(a \tens b)^\oplus=b^\oplus \tens a^\oplus\,.
\ea
\ee

We see therefore that we obtained two supersymmetric extensions of
$\kappa$-Poincar{\'e} algebra written in bicrossproduct basis. In the following
section we shall describe the graded bicrossproduct
structure.

\setcounter{equation}0
\section{Graded bicrossproduct structure of \kdef \poin superalgebra}

Let us write the classical $N=1$ \poin superalgebra as the following graded
semidirect product
\bel{3.1}
\sP = \so \bowti \o T_{4;2}\,,
\ee
where the superalgebras $S\so$ and $\o T_{4;2}$ are given respectively
by the formulae (1.3), (1.4a) and (1.4b). The cross relations describing
$\so$ covariance are
\bl
\beq
{}[M_{\m\n},P_\ro]&=&i (g_{\n\ro}P_\m - g_{\m\ro}P_\n)\,,\\[2mm]
{}[Q_\a, P_\ro]&=&0\,,\\[2mm]
{}[M_{\m\n},\o Q_\ka]&=&\frac i 2  (\s_{\m\n})_\ka{}^\kb \o Q_\kb
\\[2mm]
{}\{Q_\a,\o Q_\kb\}&=& 2(\s_\m)_{\a\kb}P^\m\,.
\eeq
\el
We see that the basic superalgebra relation (3.2d) occurs as a
covariance relation.

The graded semidirect product formula (3.1) can also be written as
follows:
\bel{3.3}
\sP = \o{\so} \bowti T_{4;2}\,,
\ee
where
$\o{\so}=(M_{\m\n},\o Q_\ka)$ and $T_{4;2}=(P_\m,Q_\a)$.
It is easy to see that the relations (3.2a-3.2d) also describe the
cross relations for \r{3.3}, with the role of $Q_\a$ and $\o Q_\ka$ in
the bicrossproduct interchanged.

The \kdef \poin superalgebras (2.35) can be treated as $\k$-deformations
of the semidirect products (3.1), taking the form
of the bicrossproducts\footnote{
The bicrossproducts of Hopf algebras were introduced by Majid
([8]; see also [10]).
The notion of crossproducts for braided quantum groups, which
can be considered the generalization of the notion of quantum
supergroups, was discussed in [11].
}. We obtain
\bl
\bel{3.4a}
\U^{(+)}_\k(\sP) = T_{4;2}^{\k(+)} \mathop{\ctb} 
O(1,3;2)
\,,
\ee
\bel{3.4b}
\U_\k^{(-)}(\sP) = O(1,3;2) \mathop{\btc}
T_{4;2}^{\k(-)}\,,
\ee
\el
where $O(1,3;2) = (M^{(0)}_{\m\n}, Q_\a^{(0)})$
and
$T_{4;2}^{\k(\pm)} = (P_\m, \o Q_\ka^{(\pm)})$.
We see from the relation (2.23) and (2.30) -- (2.33) that the Hopf
superalgebras $T_{4;2}^{\k(\pm)}$ are described by the classical
superalgebra and noncocommutative coproducts; the Hopf superalgebra
$O(1,3;2)$ is classical in the algebra as well as the
coalgebra sectors. The bicrossproduct structure of (3.4a-b)
is described by
\ben
\bl
\item[i)] The actions $\ha\pm$
\beq
\ha+ \: \TK+ \tens \so \to{} \TK+\,, \\[2mm]
\ha- \: \so \tens \TK-  \to{} \TK-\,,
\eeq
modifying the cross relations (3.2a-3.2d),
\el
\bl
\item[ii)] The coactions $\hb\pm$
\beq
\hb+ \: \so \to{} \TK+ \tens \so\,, \\[2mm]
\hb- \: \so \to{} \so  \tens \TK- \,,
\eeq
\el
modifying the classical coproducts for the generators of $\so$.
\een
Further we shall consider only the bicrossproduct described by the
action $\ha+$ and coaction $\hb+$. From the formulae (2.22c) and
(2.27-2.29) it is easy to check that $\ha+$ has the following nonlinear
(i.e.\ \kdef) components\footnote{The components which are not deformed
describe standard semidirect product.}
\bel{3.7}
\ba{rcl}
\ha+\(\P+_j\tens N_i^{(+)} \) &=& -i\delta_{ij} \left[ \frac{\k}{2} (1-
e^{-\frac{2P_0 }{\k}}) + \frac 1 {2\k} {\vec P{}^{(+)}}^2 \right]+
\frac{i}\k{\P+_i\P+_j}\,, \\[2mm]
\ha+\( \o Q^{(+)}_\ka \tens N_i^{(+)}\)& =&  \frac{i}2 e^{-\frac{P_0}\k} ( \o
Q^{(+)} \sigma _i)_\ka- \frac{1}{2\k} \epsilon_{ikl}\P+_k (\o
Q^{(+)}\s_i)_\ka\,,
\\[2mm]
\ha+\( \o Q_\kb^{(+)}\tens  Q_\a^{(+)} \) &=&4\k \delta_{\a\kb}
\sinh \frac{P_0}{2\k} - 2 e^{\frac{P_0}{2\k}} \P+_i(\sigma_i)_{\a\kb}\,.
\ea
\ee
Similarly, it follows from the formulae (2.20b) and (2.30 -- 2.33) that
the coaction $\hb+$ has the following $\k$-dependent
components\footnote{In the classical case $\hb+(I^{(0)}_A)=1 \tens
I^{(0)}_A$.}
\bel{3.8}
\ba{rcl}
\hb+\,\(N_i^{(+)}\) &=& e^{-\frac{P_0}{\k}} \tens \Np_i+\frac1\k
\e_{ijk}\P+_j\tens
M_k +\frac i {4\k} (\s_i)_{\a\kb}\epk- \o \Qp_\ka \tens \Qp_\b\,, \\[2mm]
\hb+\,\(\Qp_\a\) &=&e^{-\frac{P_0}{2\k}} \tens \Qp_\a\,.
\ea
\ee
It can be checked that the actions (3.7) and coactions (3.8) satisfy the
axioms required by the bicrossproducts structure \cite{9,10}.

The action $\ha-$ and coaction $\hb-$ can be obtained from
$\ha+$ and $\hb+$  by the
nonstandard involution (2.36), changing the sign of the parameter $\k$
as well as
the order in the tensor product.

Finally one can show that modifying properly the definitions (2.17) --
(2.18) of the bicrossproduct basis one can introduce the quantum $N=1$
\poin superalgebra as the \kdef crossproduct (3.3) supplemented by
sutably deformed crosscoproduct. In this way we obtain two other ways of
expressing quantum deformation of $\sP$ in the bicrossproduct form.

\setcounter{equation}0
\section{\kdef covariant chiral superspace}

Following the general theory of bicrossproducts [8, 10, 11]
from the formula (3.4) follows that the quantum \kdef $N=1$ supergroup
$(\sPk)^*$ dual to the \kdef \poin superalgebra (3.4a-b) can be written
also in the
bicrossproduct form (see e.g.\ \r{1.6}) \footnote{Further we shall use the
bicrossproduct from Sect.\ 3 with upper index $(+)$, and subsequently
drop this index.},
where
\begin{itemize}
\item $(\so)^*=C(SO(1,3;2))$ is the graded commutative algebra of
functions on the graded Lorentz group $SO(1,3;2)$.
\item $( T^{\k(\pm)}_{4;2})^*$ is the graded algebra of functions on
\kdef chiral superspace, dual to the
Hopf algebras $ T^{\k(\pm)}_{4;2}$.
\end{itemize}

The action and coaction  which describes the bicrossproduct Hopf
structure of the deformed graded algebra of functions on $N=1$ \poin
supergroup can be obtained from the actions and coactions (3.7 -- 3.8)
by respective dualizations.

Let us describe firstly the \kdef chiral superspace. Using the relations
\bel{4.1}
\ba{rcl}
\pair{t}{zz'}& =&(-1)^{\eta(z)\eta(t_{(2)})} \pair{t_{(1)}}z
\pair{t_{(2)}}{z'}\,, \\[2mm]
\pair{tt'}{z}&=&(-1)^{\eta(z_{(1)})\eta(t')}\pair{t}{z_{(1)}}\pair{t'}{z_{(2)}}\,,
\ea
\ee
where $t,t' \in  \sT$ (generators $P_\m$, $\o Q_\a$) and $z,z'\in \o \sT$
(generators $x^\m$, $\o \theta_\a$) and the orthonormal basis
\bel{4.2}
\ba{l}
\pair{P_\m}{x^\n}=\delta_\m^\n\,,\qquad \pair{\o Q_\ka}{\o \theta^\kb} =
i\delta_\ka^\kb\,,\\
\pair{\o Q_\ka}{x^\n} = \pair{P_\m}{\o \theta^\kb} =0\,,
\ea
\ee
one can derive that
\bel{4.3}
\ba{rl rl}
[x^0,x^k]&= - \frac 1 \k x^k\,,& [x^k,x^l]&=0\,,\\[2mm]
{}[x^0,\o\theta^\ka]&=- \frac 1 {2\k} \o \theta ^\ka\,, &\qquad [x^\m,\o
\theta^\ka] &= \{\o \theta ^\ka, \o \theta^\kb\}=0\,,
\ea
\ee
\bel{4.4}
\ba{l}
\cop x_\m = \ccop{x_\m}\,,\\[2mm]
\cop \o \theta ^\ka = \ccop{ \o \theta ^\ka}\,.
\ea
\ee

The relations (4.3 -- 4.4) describe the chiral extension of
$\k$-Minkowski space, introduced firstly  by Zakrzewski [12]. The
general formula describing the duality pairing can be written as follows
(compare with  \cite{6}).
\bel{4.5}
\ba{l}
\pair{f(P_i, P_0, \o \theta_\kb)}{:\psi(x^\m, x^0, \o \theta
^\a):}=\\[2mm]
= f\(\po{x^i},\po{x^0},i \po{\o \theta ^\b}\)
\psi(x^\m, x^0, \o \theta ^\a)_{\dsp|_{x^\m=0 \atop \o \theta ^\ka =0}}
\ea
\ee
where the normal ordering describes the functions of the $\k$-superspace
coordinates with the generators $x_0$ staying to the right from the
generators $x_i$. The canonical action of $\o \sT$ on the \kdef chiral
superspace is given by the well-known formula
\bel{4.6}
t \trr z = \pair{t}{z_{(1)}} z_{(2)}
\ee
which gives $P_\mu \trr x^\nu =\delta_\mu{}^\nu$ and
\bel{4.7}
\ba{rcl}
P_i \trr :\psi (x^i, x^0, \o \theta ^\ka): &=& :\po{x_i}\psi (x^i, x^0, \o
\theta ^\ka):\,,\\[2mm]
P_0 \trr :\psi (x^i, x^0, \o \theta ^\ka): &=& :\po{x_0}\psi (x^i, x^0, \o
\theta ^\ka):\,,\\[2mm]
\o Q_\ka \trr :\psi (x^i, x^0, \o \theta ^\ka): &=& i:\po{\o \theta
^\ka}\psi (x^i, x^0, \o \theta ^\ka):\,.
\ea
\ee

In order to describe the action of $\U(\so)$ on \kdef superspace we
dualize the action $\a$ of $\U(\so)$ on Hopf superalgebra $\o \sT$ (see
(3.7)) by the well known formula:
\bel{4.8}
\pair t {h \trr z} = \pair{\da(h\tens t)}{z}\,.
\ee

We obtain the following covariant action of the generators of $\U(\so)$ on
the \kdef chiral superspace
\bel{4.9}
\ba{rl rl rl}
M_i \trr x^0& =0\,, &\quad M_i \trr x^j&=i\epsilon_{ijk}x^k\,, &\quad M_i \trr
\o \theta ^\kb &= - \frac12 (\s_i)^\kb_\gamma\o
\theta^\gamma\,,\\[2mm]
N_i \trr x^0& =-ix^i\,, & N_i \trr x^j&=-i\delta_i{}^jx^0\,,& N_i \trr \o
\theta ^\kb &= \frac12 (\s_i)^\kb{}_\gamma \o \theta
^\gamma\,,\\[2mm]
Q_\a \trr x^0&= -2i\o \theta^\ka\,, & Q_\a \trr x^j&= 2i(\s_j)_{\a\kb}\o\theta
^\kb\,,&
Q_\a \trr \o \theta^\kb&=0\,,\\[2mm]
\ea
\ee
identical with the classical $\k$-independent
action on covariant chiral superspace
[13].
Further using the relation
\bel{4.10}
x \trr (zz')=(-1)^{\eta(z)\eta(x_{(2)})} (x_{(1)} \trr z)(x_{(2)} \trr
z')\,,
\ee
where $x \in \U_\k({\cal P}_{4;1})$, one obtains the action of the
$\k$-deformed $N=1$ \poin superalgebra generators on the functions of
the \kdef chiral superspace coordinates, It can be checked that e.g.\ the
action on the quadratic polynomials of $(x^k,\o \theta ^\a)$ contains
anomalous $\k$-dependent terms.

The differential calculus on the \kdef chiral superspace and the
theory of \kdef superfields will be described in our further
publications. Such a calculus is described by the supersymmetric
extension of the differential calculus on $\k$-Minkowski space (see [14,
15]) and it is different from the one described in [16] for quantum
superspace with the generators satisfying quadratic algebra.

\subsection*{Acknowledgements}
One of the authors (J.L.) would like to thank Prof.\ L.\ Dabrowski and
Prof. C.\ Reina for their hospitality at SISSA (Trieste), where the
paper has been completed.


\begin{thebibliography}{99}
\bibitem{1} J. Lukierski, A. Nowicki, H. Ruegg and V. Tolstoy, {\it
Phys. Lett.} {\bf B264}, 331 (1991)
\bibitem{2} S. Giller, J. Kunz, P. Kosi{\'n}ski, M. Majewski and P.
Ma{\'s}lanka, {\it Phys. Lett.} {\bf B286}, 57 (1992)
\bibitem{3} J. Lukierski, A. Nowicki and H. Ruegg, {\it Phys. Lett.}
{\bf B293}, 344 (1992)
\bibitem{4} J. Lukierski, A. Nowicki and J. Sobczyk, {\it J. Phys.}
{\bf A26}, L1109 (1993)
\bibitem{5} J. Lukierski, H. Ruegg and W. R{\"u}hl, {\it Phys.
Lett.} {\bf B313}, 357 (1993)
\bibitem{6} S. Majid and H. Ruegg, {\it Phys. Lett.} {\bf B334}, 348
(1994)
\bibitem{7} P. Kosi{\'n}ski, J. Lukierski, P. Ma{\'s}lanka and J.
Sobczyk, Wroc{\l}aw University preprint IFTUWr 868/94, March 1994; {\it
J. Phys. A}, in press
\bibitem{8} S.\ Majid, Journ.\ Alg. {\bf 130}, 17 (1990)
\bibitem{8a} J.\ Lukierski, A.\ Nowicki and H.\ Ruegg, Phys.\ Lett.\
{\bf B271}, 321 (1991)
\bibitem{9} S.\ Majid, {\em Foundations of Quantum Groups}, Chapt.\ 6,
Cambridge University Press, in press
\bibitem{10} S.\ Majid, Journ.\ Alg.\ {\bf 163}, 165 (1994)
\bibitem{11} S.\ Zakrzewski, Journ.\ of Phys, {\bf A27}, 2075 (1994)
\bibitem{12} S.\ Mandelstam, Phys.\ Lett.\ {\bf 121B}, 30 (1983)
\bibitem{14} A.\ Sitarz, {\em Noncommutative differential calculus on
$\k$-Minkowski space}, DAMTP Cambridge University preprint, September
1994
\bibitem{15} S.\ Majid and H.\ Ruegg, in preparation
\bibitem{16} T.\ Kobayashi and T.\ Uematsu, Kyoto University preprint
KUCP-47, May 1992.
\end{thebibliography}
\end{document}